
\input format.sty
\hsize 5in
\vsize 7.8in
\font\eightrm=cmr8
\font\eightti=cmti8
\font\ninerm=cmr9
\font\tenrm=cmr10

\vglue 2 cm
\centerline{\tenrm{\bf {A NONCOMMUTATIVE EXTENSION OF GRAVITY}}}

\vskip 1cm
\centerline{\tenrm { {J. MADORE and J. MOURAD}}}
\centerline {\eightti {Laboratoire de Physique Th\'eorique et Hautes En\'ergies
\footnote*{Laboratoire associ\'e au CNRS}}}
\centerline{\eightti {Universit\'e de Paris-Sud, bat. 211, 91405 Orsay,
France}}
\vskip 2 cm
\centerline {\tenrm {ABSTRACT} }

{\eightrm The commutative algebra of functions on a manifold is extended to
a noncommutative algebra by considering its tensor product with the
algebra of $n \times n$ complex matrices. Noncommutative geometry is used to
formulate
an extension of the Einstein-Hilbert action. The result is shown to be
equivalent to
the usual Kaluza-Klein theory with the manifold $SU_n$ as an internal space,
in a truncated approximation.}

\beginsection{\tenrm {\bf {1. Introduction}}}

\tenrm The breakdown of classical geometrical concepts
at very small  length scales
is an appealing idea. One of the recent alternative constructions
which could replace the usual geometrical objects, such as points and
forms, is noncommutative geometry (NCG).$^1$ The basic ingredient of
NCG is an associative  algebra. It replaces the algebra of functions on the
manifold of
classical geometry. The disappearence of points at small scales
 could offer a natural regularization
of field theory.
This program has not yet been worked out. The difficulty lies essentially
in finding the correct class of algebras  which reduce at large scales
to the commutative algebra of complex functions on a manifold, and
in having a dynamical evolution of what replaces the metrical
properties of the manifold. The latter concept has not been clearly
found in the context of NCG.

In this work we report some progress on the idea of replacing
 of a homogeneous manifold, which
plays the role of the internal space of Kaluza-Klein theories,
by the algebra of $n \times n$ complex matrices.$^3$ Stated more accurately,
it is rather the algebra of functions on the manifold that is
replaced by the noncommutative
algebra of matrices. The latter has a finite basis giving
rise to a finite number
of degrees of freedom. These replace the infinite tower of exitations
of the traditional Kaluza-Klein approach.

In Section 2 we recall the construction of the matrix geometry of
Dubois-Violette ${\it et \ al.}^2$
An Einstein-Hilbert action for the non-commutative extension of space-time
is written in Section 3, and is shown to be identical with a truncated
approximation of Kaluza-Klein theories.$^4$

\beginsection {\tenrm {\bf {2. Matrix Geometry}}}

{\tenrm A convenient basis of $M_n(\bbbc)$ consists of the identity and the
$(n^2-1)$ generators
of $SU_n$. They obey the following commutation relations
$$[\lambda_a,\lambda_b]=C^c{}_{ab}\lambda_{c} .\eqno (2.1)$$
The $C^c{}_{ab}$ are the structure constants of $SU_n$.
The analogues of vector fields are the derivations of the algebra.
These are linear maps of the algebra onto itself which obey
the Leibniz rule. They are in a one to one correspondance with the
trace-free elements of $M_{n}(\bbbc)$:
$${X_{g}}(f)=[g,f].\eqno (2.2)$$
Here ${X_{g}}$ is the derivation, $g$ and $f$ are matrices. Contrary
to the commutative case the derivations of $M_{n}(\bbbc)$ do not
form a module. A derivation multiplied by a matrix is not a derivation. A basis
of the vector space of derivations is given by $e_a$, defined by
$$ e_{a}(f)=[\lambda_{a},f].\eqno (2.3)$$
The 1-forms, elements of $\Omega ^{1}(M_{n})$,
are defined to be linear maps from
the derivations to the
algebra. A convenient basis of  1-forms is given by the
duals to $e_a$
$$\theta^{a}(e_{b})=\delta ^{a}_{b}.\eqno (2.4)$$
The wedge product of forms and the exterior derivative are
 defined as in the commutative case.
For example, if $f$ is a matrix its exterior derivative is
defined by
$$ df ( {X_{g}})={X_{g}}(f)=[g,f].\eqno (2.5)$$
It can be written as
$$df=[\lambda_{a},f]\theta^{a}.\eqno (2.6)$$
Similarly if $\omega$ is a 1-form its exterior derivative is defined by
$$d\omega (X_{g},X_{h})={X}_{g}(\omega ({X}_{h}))
-{X}_{h}(\omega ({X}_{g}))-\omega([{X}_{g},{X}_{h}]).\eqno (2.7)
$$

{}From this one may deduce the exterior derivatives of
the basis $\theta ^{a}$
$$d\theta^{a}=-{1 \over 2}C^a{}_{bc}\theta ^{b} \theta ^{c}.\eqno (2.8)$$
The product on the right hand side is  the wedge product of forms.

The integral of an $n^{2}-1$ form may be defined using the trace
$$\int f\theta^{1}\dots \theta^{n^{2}-1}={\rm Tr}(f),$$
where $f$ is a matrix.

This completes the basic notions of noncommutative differential calculus
that  we will need in the following.}

\beginsection {\tenrm {\bf {3. Algebraic Kaluza-Klein Theory}}}

{\tenrm We will define an Einstein-Hilbert action on the algebra
$${\cal A}={\cal C}(M) \otimes M_{n}(\bbbc),\eqno (3.1)$$
where ${\cal C}(M)$ is the algebra of functions on the manifold $M$.

The basic object is a metric defined as a symmetric, non-degenerate
real-valued 2-form.
In order to define such a metric we need
to find a basis of 1-forms over $\cal A$. It was shown
that $^6$
$$ \Omega ^{1}({\cal A})={\cal C}(M) \otimes \Omega ^{1}(M_{n}(\bbbc)
\oplus \Omega ^{1}(M) \otimes M_{n}(\bbbc).\eqno (3.2)$$
A basis of 1-forms on $\cal A$ consists of four
independent 1-forms over the manifold $M$, $\theta^{\alpha}$, and
the $n^{2}-1$ forms, $\theta ^{a}$, defined in the previous section.
 We shall label this basis by the index $i, 0\leq i\leq 3+n^{2}-1$.

 Next we define the moving frame $\tilde \theta^{i}$ by
 $$ \tilde \theta^{\alpha}=\theta^{\alpha}, \ \ \tilde \theta ^{r}=
 \chi ^{r}_{a}(\theta^{a}-A^{a}),\eqno (3.3)$$
 where $A^{a}$ ($\chi^{r}_{a}$) are 1-forms (functions) over $M$.
 This choice of the moving
 frame is motivated in detail elsewhere.$^4$

 The spin connection and the curvature are defined
 by means of the Cartan structure equations
 $$ 0=d\tilde \theta ^{i}+{\tilde \omega ^{i}_{j}}\wedge{\tilde \theta ^{j}},
 \eqno (3.4a) $$
 $$ \tilde \Omega ^{i}_{j}=d {\tilde \omega}^{i}_{j}+{\tilde \omega}^{i}_{k}
 \wedge{\tilde \omega}^{k}_{j}.\eqno (3.4b)$$
 The exterior derivative
 of the forms $\theta ^{a}$ was defined in eq. (2.8).

 Finally the Einstein-Hilbert action is given by
 $$ S={{\mu}^{n^{2}+1} \over {n^{2}+1}}
 \int \tilde \Omega ^{i_{0} i_{1}} \tilde \theta ^{i_2} \dots
 \tilde \theta ^{i_{n^{2}+2}} \epsilon_{i_{0} \dots i_{n^{2}+2}},\eqno (3.5)$$
 where the indices are raised with the Minkowski metric in dimension
$4+n^{2}-1$
 and $\mu$ is a mass scale. Note that the
 integral contains a trace over the matrix degrees of freedom.

 A long but straightforward calculation  gives
 the traditionnal Kaluza-Klein action in a truncated approximation,$^4$
 with the manifold $ SU_n $ as an internal space. The spectrum
 consists of a $SU_n$ gauge particle, a multiplet of adjoint
 Higgs and the graviton.
This is not very surprising since our ansatz (3.3) is analoguous
 to the one used, for instance by Cho [5], in the
 isometric approach to Kaluza-Klein theories. We have only to identify the
 basis of one forms over the the matrix algebra with the Maurer-Cartan
 forms over $SU_n$.
 However our approach does not require any approximation. We no longer
 have an infinite tower of massive particles which we neglect. The
 internal structure does not add dimensions to ordinary
 space-time as is the case
 in the traditional approach to Kaluza-Klein theories. The internal
 curvature, which contributes to an effective cosmological
 constant, may now take arbitrary values. In the traditional
 approach this value is required to be large in order for the
 internal space to be small and unobservable.
  A cosmological constant is added by hand
 to the original lagrangian in order to compensate the one resulting
 from the small internal radius.}

 \beginsection {\tenrm {\bf {4. Conclusion}}}

{\tenrm The algebra of functions on the internal manifold
of Kaluza-Klein theories has been replaced
by a matrix algebra. The extension of the gravitational
action has been formulated. The spectrum of particles is finite, and
the cosmological constant may take small values. Note that if it is too
small the gauge coupling constant becomes negligible. The
coupling to fermions has been studied elsewhere.$^4$}

\beginsection{\tenrm {\bf Acknowledgements}}

 {\tenrm  This research was partially subsidized by the CEC
Science Project No. SCI-CT91-0729.}

\beginsection{\tenrm {References}}

{ \ninerm
\item {1.} A. Connes, G\'eometrie noncommutative, InterEditions, Paris (1990).

\item {2.} M. Dubois-Violette, R. Kerner, J. Madore, J. Math. Phys.
 31 (1990) 316.

 \item {3.} J. Madore, Phys. Rev. D41 (1990) 3709.

 \item {4.} J. Madore, J. Mourad, to be published in Class. and Quant.
Gravity .

 \item {5.} Y.M. Cho, Phys. Rev. D35 (1987) 2628.

 \item {6.} M. Dubois-Violette, R. Kerner, J. Madore, J. Math. Phys. 31
 (1990) 323.}

 \end